# Enhanced Welding Operator Quality Performance Measurement: Work Experience-Integrated Bayesian Prior Determination

Yitong Li, S.M.ASCE,[1] Wenying Ji, A.M.ASCE,[2] and
Simaan M. AbouRizk, M.ASCE[3]

[1]Ph.D. Student, Department of Civil, Environmental, and Infrastructure Engineering, George Mason University, Fairfax, VA, 22030; e-mail: yli63@gmu.edu
[2]Assistant Professor, Department of Civil, Environmental, and Infrastructure Engineering, George Mason University, Fairfax, VA, 22030; e-mail: wji2@gmu.edu
[3]Professor, Department of Civil and Environmental Engineering, University of Alberta, Edmonton, Alberta, Canada T6G 2W2; e-mail: abourizk@ualberta.ca

## ABSTRACT

Measurement of operator quality performance has been challenging in the construction fabrication industry. Among various causes, the learning effect is a significant factor, which needs to be incorporated in achieving a reliable operator quality performance analysis. This research aims to enhance a previously developed operator quality performance measurement approach by incorporating the learning effect (i.e., work experience). To achieve this goal, the Plateau learning model is selected to quantitatively represent the relationship between quality performance and work experience through a beta-binomial regression approach. Based on this relationship, an informative prior determination approach, which incorporates operator work experience information, is developed to enhance the previous Bayesian-based operator quality performance measurement. Academically, this research provides a systematic approach to derive Bayesian informative priors through integrating multi-source information. Practically, the proposed approach reliably measures operator quality performance in fabrication quality control processes.

## INTRODUCTION

Pipe spool fabrication is essential to the successful delivery of an industrial construction project. During the process of pipe fabrication, welding is an important step and its quality must be examined to ensure the specified requirements are satisfied. Although welding is undertaken by skilled operators, variations commonly exist in welding operator quality performance due to the lack of essential knowledge and skills (Ji et al. 2018). Therefore, being able to reliably measure welding operator quality performance is crucial since the reliable performance measurement leads to considerable advancement in project quality performance, which would further decrease rework cost and overcome schedule delays. To achieve this goal, Ji et al. (2019) have developed a Bayesian statistics-based method to estimate operator quality performance by assuming operator quality performance is stationary over time. However, one of the most significant factors—the learning effect (i.e., the continuously improved quality performance as operator work experience increases)—was neglected, which leads to a biased estimation of operator quality performance.

This research aims to enhance the previously developed approach to reliably measure welding operator quality performance by incorporating the effect of work experience. Specifically, the objective is achieved by (1) selecting a learning curve model to describe the relationship between quality performance and work experience; (2) applying the beta-binomial regression to derive the equation of the selected model; (3) determining a informative prior to represent quality performance for a given operator; (4) demonstrating the advantages of the enhanced Bayesian-based approach using a case study. The remainder of this paper is arranged as follows. In the next section, previous work on Bayesian-based operator quality performance measurement is discussed. After that, the newly proposed methodology is introduced step by step. In the following section, a practical case study is conducted to demonstrate the advantages of the newly proposed approach. Finally, contributions, limitations, and future work are concluded.

**PREVIOUS WORK**

Previously, Ji and AbouRizk (2017) have advocated the advantages of using a Bayesian-based operator quality performance measurement to incorporate inspection sampling uncertainty. In their research, operator quality performance is reflected by fraction nonconforming (i.e., percentage repair rate) as indicated below:

$$p = \frac{X}{n} \quad (1)$$

Where $p$ denotes fraction nonconforming, $X$ denotes the number of welds which fails inspections, and $n$ denotes the total number of welds. To cover the sampling uncertainty, a beta distribution $Beta(a, b)$ was chosen to model the prior distribution for the Bayesian-based fraction nonconforming estimation. The prior distribution represents operator quality performance when no inspection results are collected. The posterior distribution describes the latest measurement of operator quality performance by continuously adding more inspection results. An analytical solution for the posterior distribution follows:

$$Beta(X + a, n - X + b) \quad (2)$$

In Bayesian statistics, two types of priors are commonly used, namely, informative priors—probability distributions derived from historical data or subjective knowledge; and noninformative priors—vague, flat, and diffuse probability distributions that have the lowest bias to prior estimation when information is insufficient (Ji and AbouRizk 2017).

In estimating the welding operator quality performance, Ji and AbouRizk (2017) used a noninformative prior distribution $Beta(1/2, 1/2)$ without incorporating the learning effect of operators. The reliability of a Bayesian statistic-based method is heavily dependent on the prior determination (Winkler 1967). Incapable of determining reliable priors leads to unreliable posterior inferences, which further misleads practical decision support. Therefore, in aims of improving the existing approach, an informative prior determination method, which is able to incorporate work experience, is developed in this study.

**METHODOLOGY**

The research methodology of this study is demonstrated as Figure 1. First, the Plateau learning curve model is selected to illustrate the relationship between operator quality performance and work experience. Then, a beta-binomial regression approach is utilized to derive the unknown parameters for the selected learning curve model. After that, informative priors are determined through the derived learning curve equation. Lastly, posterior distributions are computed by incorporating newly collected inspection data.

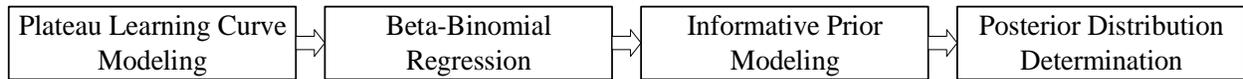

**Figure 1. Research methodology flow chart.**

**Plateau Learning Curve Modeling.** The Plateau model (Baloff 1971) describes a linear-log model with a constant term which indicates the operator's steady-state performance. The Plateau model is selected to represent the relation between welding operator quality performance and work experience. It is applicable in this research because operator quality performance reaches a steady-state as operators gain enough practices. The Plateau model in this research is represented in Equation (3):

$$FN = A + B(n)^{-C} \qquad (3)$$

Where FN denotes fraction nonconforming and $n$ denotes the total number of welds. A, B, and C are unknown parameters that can be derived using the regression approach described in the next section.

**Beta-Binomial Regression.** Regression is a statistical technique to determine the relationship between dependent variable and independent variables. For this research, the beta-binomial regression model is selected to derive parameters in the Plateau model.

R's *gamlss* package (Stasinopoulos and Rigby 2018) is a regression package which allows all the parameters of the distribution of the dependent variable to be modeled as non-linear functions of the independent variables (Rigby and Stasinopoulos 2005). In this study, *gamlss* function is used to model the mean and the variation of fraction nonconforming (i.e., dependent variable) as a non-linear function of the total number of welds (i.e., independent variable). Here, variations of fraction nonconforming are assumed to be the same for all values of total welds and are represented as $\sigma_{FN}$. The relationship between the mean value of fraction nonconforming and total number of welds follows Equation (3) can be represented as:

$$\mu_{FN} = A + B(n)^{-C} \qquad (4)$$

This equation allows defining an exclusive mean value of fraction nonconforming for every operator based on their total number of welds (i.e., work experience).

**Informative Prior Modeling.** In the Bayesian-based operator quality performance measurement approach, the prior distribution of fraction nonconforming is represented with a beta distribution as shown in Equation (5), which can be reparametrized using $\mu$ and $\sigma$, where $\mu$ (shown as Equation (6)) is the mean value of a beta distribution, and $\sigma$ (shown as Equation (7)) represents the spread of the distribution. The reparametrized equation is shown as Equation (8).

$$Beta(a, b) \tag{5}$$

$$\mu = \frac{a}{a+b} \tag{6}$$

$$\sigma = \frac{1}{a+b} \tag{7}$$

$$Beta\left(\frac{\mu_{FN,i}}{\sigma_{FN}}, \frac{1 - \mu_{FN,i}}{\sigma_{FN}}\right) \tag{8}$$

In Equation (8), $\mu_{FN,i}$ is computed from Equation (4) which represents the mean value of fraction nonconforming for operator $i$ with the total number of welds $n_i$. The reparametrized beta distribution is used as the informative prior distribution for the Bayesian-based approach.

**Posterior Distribution Determination.** After obtaining the prior distribution, the posterior distribution can be computed by following the same process as discussed in Equation (2).

**CASE STUDY**

**Data Source.** The same dataset that contains information on engineering design system and quality management system from an industrial pipe fabrication company in Edmonton, Canada is used (Ji et al. 2019). The engineering design system stores information of pipe design attributes which are grouped in the format of (nominal pipe size, pipe schedule, material type, weld type) to represent a type of welds. The quality management system stores inspection records for various pipe types, inspection records for a specified weld type which can further be summarized as inspection results shown in Table 1. The detailed description of the dataset and data processing steps are referred to the authors' previous research (Ji et al. 2019).

**Table 1. Sample inspection results for a specified weld type.**

| Operator ID | Total number of welds | Repaired welds | Fraction nonconforming |
|---|---|---|---|
| 1 | 208 | 9 | 0.043 |
| 2 | 141 | 5 | 0.035 |
| … | … | … | … |

To verify the learning effect exists in the studied dataset, relationships between fraction nonconforming and total number of welds for four common types of welds are shown as scatter plots in Figure 2. All these plots demonstrate improved quality performance with increased work experience, which also proofs the motivation of this study.

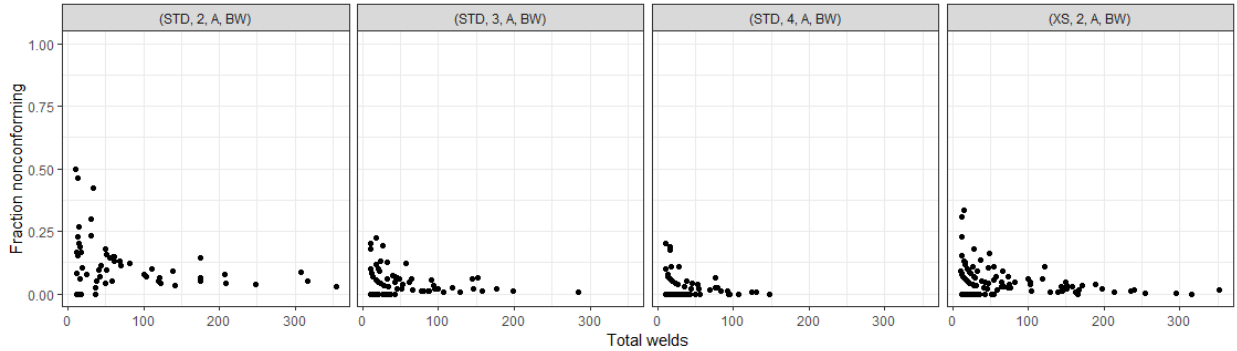

**Figure 2. Relationship between fraction nonconforming and total manufactured welds for four commonly used weld types.**

**Main Outputs.** To demonstrate the feasibility of the proposed methodology, the weld type with design attributes (STD, 2, A, BW) is selected and further analyzed since it is the most common weld type in the studied dataset. The processed data indicates that 57 welding operators had experience in producing this weld type. In Figure 3, the line is the fitted Plateau learning curve using the beta-binomial regression. The value of $\sigma_{FN}$ equals to $0.184 \times 10^{-1}$ and the computed Plateau learning curve follows the equation:

$$\mu_{FN,i} = 0.149 - 0.544 \times 10^{-2} \times (n_i)^{0.5} \tag{9}$$

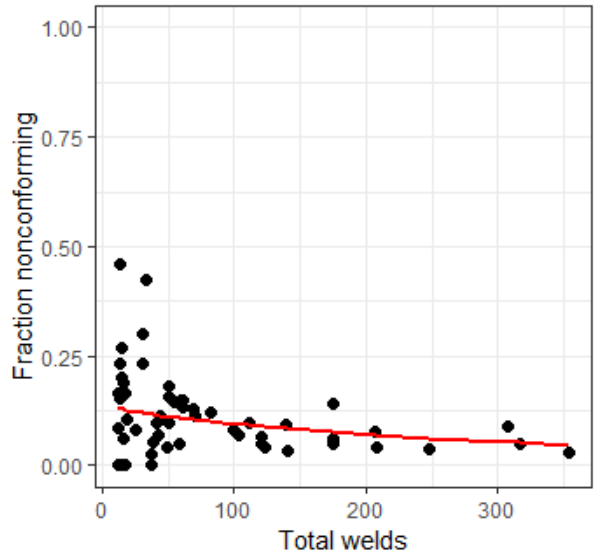

**Figure 3. Relationship between fraction nonconforming and total number of welds for weld type (STD, 2, A, BW).**

Using Equation (9), the mean value of fraction nonconforming for a welding operator can be computed based on his work experience (i.e., total number of welds). Table 2 shows the sample calculations for welding operators, which includes welding inspection information, mean value of fraction nonconforming (as per Equation (9)), informative prior distribution (as per Equation (8)), posterior distribution (as per Equation (2)), and mean value of fraction nonconforming computed from the posterior distributions (use Equation (6)).

**Table 2. Bayesian-statistics using informative prior determination for weld type (STD, 2, A, BW).**

| Operator ID | n | X | $\mu_{FN}$ | Prior: $Beta(a,b)$ | Posterior: $Beta(X+a, n-X+b)$ | Posterior mean |
|---|---|---|---|---|---|---|
| 1 | 175 | 25 | 0.077 | 4.172, 50.005 | 29.172, 200.005 | 0.127 |
| 2 | 111 | 11 | 0.092 | 4.966, 49.211 | 15.966, 149.211 | 0.097 |
| … | … | … | … | … | … | … |

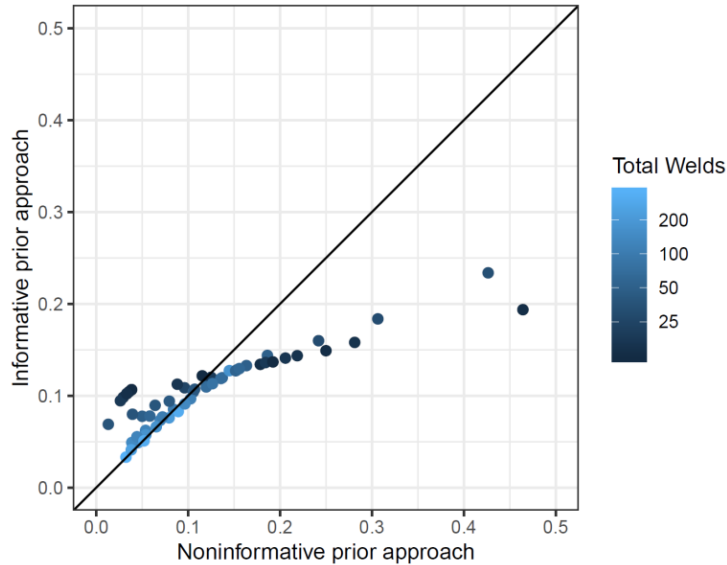

**Figure 4. Relationship of mean fraction nonconforming between informative prior approach and noninformative prior approach.**

Figure 4 illustrates differences between the proposed informative prior approach and the previous noninformative prior approach as the total number of welds changes. In this figure, the x-axis represents the mean value of fraction nonconforming computed using the noninformative prior approach and the y-axis shows the mean value of fraction nonconforming computed using the informative prior approach. The legend color represents the total number of welds (i.e., a darker color represents a low total number of welds and a lighter color represents a high total number of welds). The diagonal line represents situations when there is no difference between using the two approaches. From Figure 4, it is observed that the majority of lighter color points fall onto the diagonal line, which indicates there is no significant difference with using the two approaches when the total number of welds is high. However, deviations of the darker color points from the diagonal line indicate the differences between using the two approaches when the total number of welds is low. In summary, when the total number of welds is low, the noninformative prior approach cannot provide a quality measurement which is as reliable as the informative prior approach since the noninformative prior approach does not incorporate work experience.

In the authors' previous research, 17 welding operators, who have manufactured weld type (STD, 2, A, BW), were selected to demonstrate the feasibility of noninformative prior approach (Ji et al. 2019). For comparison purpose, the proposed informative prior determination approach is conducted on the same group of operators. Results of the two approaches are demonstrated as boxplots shown in Figure 5.

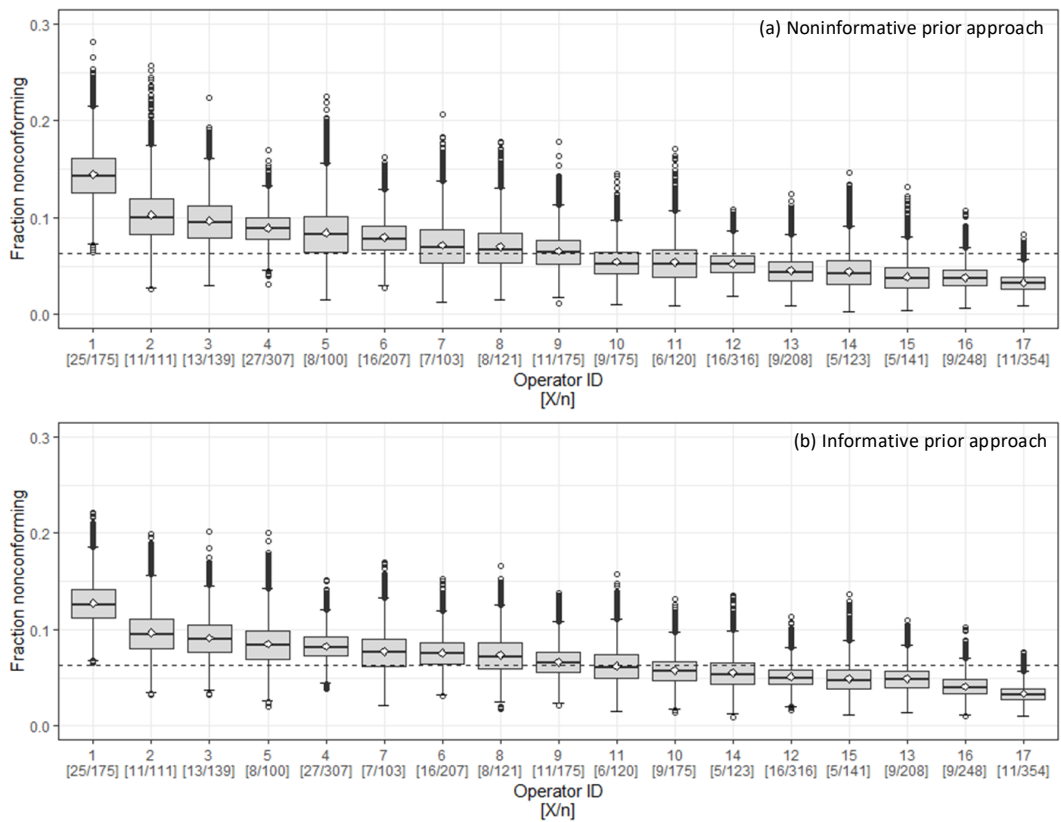

**Figure 5. Posterior distributions of fraction nonconforming for 17 welding operators: (a) noninformative prior approach and (b) informative prior approach.**

In Figure 5 (a), welding operators are ordered decreasingly as per mean values of their fraction nonconforming with ID numbers reassigned from 1 to 17. Compared to Figure 5(a), multiple changes in rankings are observed in Figure 5 (b). These changes demonstrate work experience does have an impact on welding operator quality performance, which proves that the informative prior approach is capable of measuring more realistic operator quality performance than the noninformative prior approach.

**CONCLUSION**

This research enhances the previously developed operator quality performance measurement approach to reliably measure welding operator quality performance by incorporating the effect of work experience. In this research, the Plateau learning curve model is utilized to represent the relationship between operator quality performance and work experience. A quantitative representation of the relationship is derived using the beta-binomial regression model. Based on the relationship, informative priors are determined and then utilized to obtain posterior distributions to reflect operator quality performance.

Academically, this research developed a systematic informative prior determination approach which is capable of incorporating rich information to represent the variable of interest. Practically, the proposed approach can be used to measure operator quality performance and can be used to provide practitioners with guidance in decision-making processes for improved project quality control.

Still, operator quality performance is subject to various factors (e.g. training levels and working conditions). Therefore, in the future, studies on identifying and quantifying factors that related to operator quality performance will be performed to achieve a better measurement of operator quality performance.